\documentstyle[aps,prl,amssymb]{revtex}

\input epsf

\begin{document}
\draft
\title{Calculation of mode coupling for quadrupole excitations in a BEC}
\author{G. Hechenblaikner, S.A. Morgan, E. Hodby, O.M. Marag\`o, and C.J. Foot }
\address{Clarendon Laboratory, Department of Physics, University of Oxford,\\
Parks Road, Oxford, OX1 3PU, \\
United Kingdom.}
\date{\today}
\maketitle
\begin{abstract}
In this paper we give a theoretical description of resonant
coupling between two collective excitations of a Bose condensed
gas (BEC) on, or close, to a second harmonic resonance. Using
analytic expressions for the quasi-particle wavefunctions we show
that the coupling between quadrupole modes is strong, leading to a
coupling time of a few milliseconds (for a TOP trap with radial
frequency $\sim 100\hspace{0.1em}$ Hz and $\sim 10^4$ atoms).
Using the hydrodynamic approximation, we derive analytic
expression for the coupling matrix element. These can be used with
an effective Hamiltonian (that we also derive) to describe the
dynamics of the coupling process and the associated squeezing
effects.
\end{abstract}

\section{Introduction}
In two recent experiments \cite{hechenblaikner,hodby} we observed
resonant coupling between the low-energy modes of oscillation in a
Bose-condensed gas. In the first experiment \cite{hechenblaikner}
we excited an even parity quadrupole mode (the $m=0$ low-lying
mode) and observed transfer of energy to a mode at twice the
original frequency (the $m=0$ high-lying mode). The oscillations
at the second harmonic were observed as soon as the  excitation
period ended and stayed constant in amplitude. This indicates
strong coupling between the modes so that energy  is transferred
between them at a rate comparable to the mode oscillation
frequency of a few hundred Hz, i.e. this an allowed transition
between the vibrational modes. In contrast, the coupling between a
scissors mode and a mode at half the initial frequency was found
to be a much slower process \cite{hodby}. This paper shows that
the simple downconversion process is forbidden, i.e. the matrix
element for the direct conversion of one quantum of the
higher-lying scissors mode into two quanta of a lower-lying mode,
is zero. This means that some more complicated process is required
to explain the experimental results. We also show how to calculate
the coupling rates between various modes analytically. For
resonant coupling between the quadrupole excitations we present a
simple expression for the radial integrand of the matrix element
that shows that the coupling mostly takes place in the boundary
regions of the condensate. Finally, we show that the coupling is
well described by a simple Hamiltonian which can be used for
quantitative studies of the squeezing effects related to the
harmonic generation processes.

The paper is structured as follows: Section \ref{condex} presents
the nonlinear Schr\"odinger equation (NLSE) and the derivation of
the Bogoliubov-de Gennes (BdG) equation from the many-body
Hamiltonian. These equations form the basis of the following
sections. In section \ref{hydrosolsection} we summarize the
derivation of solutions to the BdG equations in the hydrodynamic
limit following the method given in \cite{ohberg}. The assumptions
and approximations made in that derivation are important for
understanding the calculations in section
\ref{calcnonlincouprate}. Section \ref{modelsection} gives the
derivation of the Hamiltonian describing second-harmonic
generation SHG or degenerate down-conversion from the NLSE,
closely following the approach given in \cite{morgan}. The
coupling matrix elements governing the nonlinear processes are
calculated in section \ref{calcnonlincouprate}. A simple
expression is found for resonant coupling and the results are
compared to an exact numerical calculation. We show that symmetry
arguments forbid the direct down-conversion of the scissors mode
and discuss our results with respect to two recent experiments
\cite{hechenblaikner,hodby} by our group.

\section{Condensate Excitations}
\label{condex}
 Our treatment of the coupling between the modes
starts with the Gross-Pitaevskii equation (GPE) for the
macroscopic wave function $\Psi({\bf r},t)$ (also called the order
parameter):
\begin{equation}
i \hbar \frac{\partial \Psi}{\partial
t}=\left[-\frac{\hbar^2}{2m}\nabla^2+V_{ext}+g|\Psi|^2\right]\Psi
\label{GP}.
\end{equation}
The external potential for a harmonic trap is $V_{ext}({\bf r})=
m\sum_i \omega_i^2 x_i^2/2$ and $g=4\pi\hbar^2 a_s/m$
characterizes the nonlinearity which depends on the particle
interaction strength through the scattering length $a_s$. The
ground state $\Psi_g$ is the lowest energy eigenstate of the
condensate and a solution to the time-independent NLSE:

\begin{equation}
\left(-\frac{\hbar^2}{2m}\nabla^2+V_{ext}+g|\Psi_g|^2\right)\Psi_g=\mu\Psi_g.
\label{groundstateeqn}
\end{equation}
where the energy of the ground state $\mu$ the chemical potential
of the system.

 One way to derive the collective excitations is to
linearize the GPE for small perturbations around the ground state
with the ansatz
\begin{equation}
\Psi({\bf r},t)=e^{-i \mu t}\left[\Psi_g({\bf
r})+\sum_i\bigg(u_i({\bf r}) b_i e^{-i \omega_i t} + v_i^*({\bf
r})b_i^*e^{+i\omega_i t}\bigg)\right]. \label{classicansatz}
\end{equation}
Substitution into the GPE and linearisation with respect to the
small ampitudes $b_i$ yields the Bogoliubov-de Gennes (BdG)
equations
\begin{eqnarray}
{\cal L} u_i+g\Psi_g^2 v_i &=& \hbar\omega_i u_i\nonumber\\
 {\cal L}v_i+g\Psi_g^{*2}u_i &=&-\hbar\omega_i v_i.
\label{BdG}
 \end{eqnarray}
The operator ${\cal L}$ is given by
\begin{equation}
{\cal L} = -\frac{\hbar^2}{2m}\nabla^2+V_{ext}({\bf r}) - \mu
+2g|\Psi_g|^2.
\end{equation}
By solving the BdG equations we find the eigenmodes with energies
$\hbar \omega_i$, and wavefunctions $v_i,u_i$ that satisfy the
orthogonality and symmetry relations:

\begin{eqnarray}
\int d^3{\bf r} \left(u_i u_j^*-v_i v_j^*\right) &=& \delta_{ij}
\nonumber\\ \int d^3{\bf r} \left(u_i v_j^*-v_i u_j^*\right) &=&
0. \label{orthogonality}
\end{eqnarray}

The small complex amplitude coefficients $b_i,b_i^*$ in
Eq.(\ref{classicansatz}) can be replaced by annihilation and
creation operators $\hat{b}_i,\hat{b}_i^{\dag}$ respectively. This
is justified by the standard approach of second quantization,
where the eigenmodes of a classical system are found and then the
complex amplitudes are replaced by mode operators. Alternatively
one can start with the Grand Canonical many-body Hamiltonian for
the field operator $\hat{\Psi}({\bf r},t)$,

\begin{equation}
\hat{H}=\int d^3{\bf r} \hat{\Psi}^{\dag}({\bf
r},t)\left[-\frac{\hbar^2}{2m}\nabla^2+ V_{ext}({\bf r})-\mu
+\frac{g}{2} \hat{\Psi}^{\dag}({\bf r},t)\hat{\Psi}({\bf
r},t)\right]\hat{\Psi}({\bf r},t), \label{manybodyhamiltonian}
\end{equation}
and make the ansatz:
\begin{equation}
\hat{\Psi}({\bf r},t)=\Psi_g+\sum_i\left(u_i({\bf
r})\hat{b}_i(t)+v_i^*({\bf r})\hat{b}_i^{\dag}(t)\right).
\label{quantansatz}
\end{equation}

In this approach the field operator is split into its expectation
value (the condensate part) and a fluctuating part that accounts
for collective excitations and the thermal cloud. Substitution of
Eq.(\ref{quantansatz}) into the Hamiltonian of
Eq.(\ref{manybodyhamiltonian}), and neglecting terms of order
three or four in the excitation operators
$\hat{b}_i,\hat{b}_i^{\dag}$ gives a quadratic Hamiltonian which
is diagonalized exactly if $\psi_g$ satisfies the GPE of
Eq.(\ref{groundstateeqn}) and the wavefunctions $u_i,v_i$ are
solutions of the BdG Eqs.(\ref{BdG}). The Hamiltonian can
therefore be written as
\begin{equation}
\hat{H}=E_g+\sum_i \hbar\omega_i  \hat{b}_i^{\dag} \hat{b}_i+C.
\label{manybodyenergy}
\end{equation}
Here $C$ is the zero-point energy of the non-condensate and $E_g$
is the energy of the condensate given by
\begin{equation}
E_g=\int d^3{\bf r}
\Psi_g^*\left[-\frac{\hbar^2}{2m}\nabla^2+V_{ext}-\mu+\frac{g}{2}|\Psi_g|^2\right]\Psi_g.
\end{equation}
So far, in all experiments on collective excitations the
eigenmodes have been excited strongly into a coherent state. For
these conditions one can assume that the mode operators commute
and replace them by complex numbers so that Eq.(\ref{quantansatz})
reduces to Eq.(\ref{classicansatz}) except for a factor $e^{-i\mu
t}$ which amounts to a shift in the zero of energy.

\section{Calculating the quasparticle wavefunctions in the
hydrodynamic limit} \label{hydrosolsection}

In this section we give a brief overview of how to calculate the
excited state wavefunctions $u_i,v_i$ directly following the
approach given in \cite{ohberg}. The starting point are the BdG
Eqs.(\ref{BdG}). These can be rewritten in dimensionless units by
introducing the following coordinate transforms:
$y_j=r_j/l_j,j=1..3$, where $l_j=(2\mu/m\omega_j^2)^{1/2}$ are the
characteristic lengths of the condensate in the Thomas Fermi
regime. As in \cite{ohberg} we define the small dimensionless
parameter $\xi=\hbar\bar{\omega}/2\mu$ and the dimensionless
energy of mode $i$, $\epsilon_i=E_i/\hbar\bar{\omega}$, where
$E_i$ is the energy of mode $i$ and
$\bar{\omega}=(\omega_1\omega_2\omega_3)^{1/3}$ is the geometric
mean oscillator frequency. We also introduce the mean
characteristic length $l_c=(l_1 l_2 l_3)^{1/3}$ and the
dimensionless Laplace operator
$\tilde{\triangle}=\sum_{j=1}^3(\omega_j/\bar{\omega})\partial^2/\partial
y_j^2$ and define $y^2=\sum_{j=1}^3 y_j^2$. The resulting
equations are:
\begin{eqnarray}
-\xi^2\tilde{\triangle}u_i+y^2
u_i+(2u_i+v_i)\bar{n}_0=(1+2\xi\epsilon_i)u_i,\label{ohberg1}\\
-\xi^2\tilde{\triangle}v_i+y^2
v_i+(2v_i+u_i)\bar{n}_0=(1-2\xi\epsilon_i)v_i,\label{ohberg2}\\
-\xi^2\tilde{\triangle}\psi_g+y^2\psi_g+\bar{n}_0\psi_g=\psi_g,
\end{eqnarray}
where $\bar{n}_0=|\psi_g|^2g/\mu$. These equations can be combined
to form fourth order equations for the functions $f_i^{\pm}=u_i
\pm v_i$. In the hydrodynamic limit the expression
$\psi_g=\sqrt{n_0(1-y^2)}$, where $n_0=\mu/g$ is the maximum
condensate density, is then substituted for the ground state wave
function $\psi_g$, and terms of second order in the small
parameter $\xi$ are omitted.

Now we introduce the operator $\hat{G}$ with the definition:
\begin{equation}
\hat{G}=(1-y^2)\tilde{\triangle}-2\sum_ iy_i(\omega_i/
\bar{\omega})^2\partial/\partial y_i.
\end{equation}
and define new functions $W_i(y_1,y_2,y_3)$ by $f^{\pm}_i
(y)=C^{\pm}_i(1-y^2)^{\mp 1/2}W_i(y_1,y_2,y_3)$, where the
relation between the coefficients $C^{\pm}_i$ is given by
$C^+_i=\epsilon\xi C^-_i$. Using these definitions one finally
obtains from Eqs.(\ref{ohberg1},\ref{ohberg2}) the compact
expression \cite{ohberg}:
\begin{equation}
\hat{G}W+2\epsilon^2W=0, \label{Gequation}
\end{equation}
where we have omitted the mode-index $i$ for simplicity. For a
spherical trap this equation can be solved exactly. It is a
hypergeometric differential equation with Jacobi polynomials as
the general solution. The quantization of the energies comes from
the condition that the function must converge at the condensate
boundaries which yields an analytic expression for the mode
spectrum \cite{stringari-modespectrum,ohberg}.

In the most general case of an anisotropic trap with three
different trap frequencies, one can make a polynomial ansatz. The
symmetry of the Hamiltonian means that parity is a good quantum
number for any spatial coordinate. We shall find the solutions for
the quadrupolar modes where the order of the polynomial is $2$. We
can make the ansatz \cite{ohberg}
\begin{equation}
W \propto y_iy_j,\hspace{0.2em} i\neq j \label{scissorseq}
\end{equation}
 to find the three odd parity
eigenfunctions with eigenfrequencies
$\Omega=(\omega_i^2+\omega_j^2)^{1/2}$. These so-called scissors
modes have been studied extensively by our group \cite{marago} and
we use Eq.(\ref{scissorseq}) to derive some of their coupling
properties in section \ref{calcnonlincouprate}. We have to make a
different ansatz to find the three even parity eigenfunctions
(which are also referred to as diagonal quadrupolar modes
\cite{pires}):
\begin{equation}
W\propto 1+\sum_{j=1}^3 b_j(\bar{\omega}/\omega_j)^2y_j^2.
\label{polynomialW}
\end{equation}
The polynomial coefficients $b_j$ completely characterize the mode
geometry and will be very important later on in our expression for
the coupling matrix element. In the following we use the
abbreviation $\bar{b}_j= b_j(\bar{\omega}/\omega_j)^2$.
Substitution of Eq.(\ref{polynomialW}) into Eq.(\ref{Gequation})
shows that $W$ is a solution provided that the following equations
hold
\begin{eqnarray}
S\left(\begin{array}{c} b_1 \\b_2\\b_3\end{array}\right)=0,
\label{Sequations}\\ \sum_{j=1}^3
b_j+\frac{\Omega^2}{\bar{\omega}}=0 \label{normequation},
\end{eqnarray}
where the matrix $S$ is defined as
\begin{equation}
S=\left(
\begin{array}{ccc}
 3-\frac{\Omega^2}{\omega_1^2} & 1 &1 \\ 1 &
3-\frac{\Omega^2}{\omega_2^2} & 1 \\
 1 & 1 & 3-\frac{\Omega^2}{\omega_3^2}
\end{array} \right).
\end{equation}
 The eigenfrequencies $\Omega$ are found by demanding that $det(S)=0$.
 The resulting equation is
\begin{equation}
\Omega^6-3\Omega^4(\omega_1^2+\omega_2^2+\omega_3^2)+8\Omega^2(\omega_2^2
\omega_3^2+\omega_1^2\omega_3^2+\omega_1^2\omega_2^2)-20(\omega_1^2\omega_2^2\omega_3^2)=0.
\label{frequencyequation}
\end{equation}
This general expression simplifies for the case of an axially
symmetric trap ($\omega_1=\omega_2$). In this case the solutions
are $\Omega=\sqrt{2}\omega_1$ for the $m=2$ mode, and
\begin{equation}
\Omega^2=2+\frac{3}{2}\lambda^2\mp\frac{1}{2}\sqrt{9\lambda^4+16\lambda^2+16}
\label{topequation}
\end{equation}
for the $m=0$ low-lying and the $m=0$ high-lying mode. Here, the
trap anisotropy is given by $\lambda=\omega_z/\omega_r$, where
$\omega_z,\omega_r$ are the axial and radial trap frequencies
respectively. Eqs.(\ref{frequencyequation},\ref{topequation}) were
derived in the review paper on BEC  by Dalfovo et al.
\cite{dalfovo}. In the next step the polynomial coefficients $b_j$
are found from any two of the three equations in
(\ref{Sequations}) and Eq.(\ref{normequation}).

So far we have summarised important known results that enable us
to calculate the quasi-particle wavefunctions for all six
quadrupole modes. We will use these in section
\ref{calcnonlincouprate} to calculate the matrix elements which
describe the couling of these modes.

\section{A model for second harmonic coupling between two modes}
\label{modelsection} We follow the approach given in \cite{morgan}
to derive a set of coupled nonlinear equations (describing second
harmonic generation) from the NLSE.
 For convenience we normalize
the condensate wavefunction to unity and change the parameter $g$
in Eq.(\ref{GP}) to $N_0 g$, where $N_0$ is the number of
particles in the condensate. We introduce a set of excitations
that is normal to the condensate and also diagonalises the
many-body Hamiltonian of Eq. (\ref{manybodyhamiltonian}). This is
achieved by projecting out the overlap with the condensate from
the solutions to the BdG equations to give quasi-particle
wavefunctions defined by
\begin{eqnarray}
\tilde{u}_i &=& u_i-c_i\Psi_g \\ \tilde{v}_i^* & =&
v_i^*+c_i^*\Psi_g,
\end{eqnarray}
where $c_i=\int d^3{\bf r}\left[\Psi_g^* u_i\right]=-\int d^3{\bf
r}\left[\Psi_g v_i\right]$. These wavefunctions still diagonalise
the many-body Hamiltonian (\ref{manybodyhamiltonian}) and the
orthogonality relations (\ref{orthogonality}) hold as well. The
advantage of introducing excitations orthogonal to the ground
state is that it makes it easier to extract the amplitudes of
various excitations from a given wavefunction. In terms of the
orthogonal excitations a general wavefunction can be written as
\cite{morgan}:
\begin{equation}
\Psi({\bf r},t)=e^{-i\mu t}\left\{(1+b_g)\Psi_g({\bf
r})+\sum_{i>0}\left[\tilde{u}_i({\bf r})b_i(t)+\tilde{v}_i^*({\bf
r})b_i^*(t)\right]\right\}, \label{newdefinition}
\end{equation}
where the coefficient $b_g$ describes the change in the
condensate. It is easy to show that for the orthogonal excitations
the following relationships hold
\begin{eqnarray}
\int d^3{\bf r}\hspace{0.5 em} \psi_g^*\Psi e^{+i\mu t} = 1+b_g
\nonumber\\ \int d^3{\bf r} \left[ \tilde{u}_i^*\Psi e^{+i\mu
t}-\tilde{v}_i^*\Psi^*e^{-i\mu t}\right]=b_i. \label{orthoproject}
\end{eqnarray}
The population of the condensate ground state is given by
$|1+b_g|^2 N_0$ and the population of the excited states by
$|b_i|^2 N_0$.

In the next step we obtain the equations of evolution for the
complex coefficients $b_i(t)$ by substituting the expansion of the
wavefunction (\ref{newdefinition}) into the GPE (\ref{GP}), and
carrying out the projections described by Eqs.(\ref{orthoproject})
. We so obtain the Heisenberg equations for the c-number
equivalents of the mode operators $b_i$. We then transform these
equations for the mode amplitudes into the interaction picture by
making the ansatz $b_i(t)=b_i^R(t) e^{-i\omega_it}$. This gives
rise to a large number of terms oscillating at frequencies
$\omega_i\pm\omega_k\pm\omega_j$. If we focus on second harmonic
processes where $\omega_k=\omega_i$ and $\omega_j \simeq
2\omega_i$ we can neglect all the rapidly oscillating terms and
retain only the term oscillating at
$\Delta_{ij}=\omega_j-2\omega_i$. This is called the Rotating Wave
Approximation (RWA). If we neglect any variation in the population
of the condensate mode we obtain the following coupled equations
of motion for the two modes $i=1$ and $ j=2$:
\begin{eqnarray}
i\hbar\frac{d b_1^R}{dt}=N_0 g M_{12} b_1^{R*} b_2^R
e^{-i\Delta_{12}t}\\ i\hbar \frac{d b_2^R}{dt}=\frac{1}{2}N_0 g
M_{12}^* b_1^R b_1^R e^{i\Delta_{12}t}, \label{coupledbees}
\end{eqnarray}
where the matrix element $M_{12}$ is given by
\begin{equation}
M_{12}=2\int d^3{\bf
r}\left[\psi_g^*\left(2\tilde{u}_1^*\tilde{v}_1^*\tilde{u}_2+\tilde{v}_1^*\tilde{v}_1^*\tilde{v}_2\right)+
\psi_g\left(2\tilde{u}_1^*\tilde{v}_1^*\tilde{v}_2+\tilde{u}_1^*\tilde{u}_1^*\tilde{u}_2\right)\right].
\label{matrixelement}
\end{equation}

These equations describe the transfer of excitation between the
two modes via annihilation (creation) of two quanta in mode $1$
and creation (annihilation) of one quantum in mode $2$ which is
also called a second harmonic process. The matrix element $M_{12}$
contains all the information on the geometry of the two modes that
are coupled. If we excite the lower mode at resonance
($\Delta_{12}=0$) and there is no initial population in the upper
mode, then all the excitation is transferred to the upper mode.
The opposite is not true, i.e. there is no transfer from an
initially excited upper mode to the lower mode if we start off
with zero population in the lower mode. We will see later in this
section that in a quantum mechanical description, where the lower
mode is described by operators rather than c-numbers,
downconversion does occur.
 The strength
of the processes depends on the spatial overlap between the
respective quasiparticle wavefunctions. The characteristic time
scale for the transfer from mode $1$ to mode $2$ is given by the
expression:
\begin{equation}
T=\left|\frac{\sqrt{2}\hbar}{NgM_{12}b_1(0)}\right|.
\label{transfertime}
\end{equation}
This model for the second harmonic coupling between two collective
excitations allows us to find an explicit expression for the
matrix element governing the process.

\subsection{A Quantum Mechanical Model for the Coupling}
 Alternatively, the two coupled nonlinear
equations (\ref{coupledbees}) can be derived from the Hamiltonian
(\ref{secondquantham}), which gives a full quantum mechanical
description and clearly shows the underlying physical processes
\begin{equation}
H = \hbar\omega_1 \hat{a}_1^\dag \hat{a}_1+ 2\hbar\omega_1
\hat{a}_2^\dag \hat{a}_2 +\frac{\hbar\kappa}{2}(\hat{a}_1^{\dag
2}\hat{a}_2+\hat{a}_1^2 \hat{a}_2^{\dag}),\label{secondquantham}
\end{equation}
where $\hat{a}_1^{\dag}\hat{a}_1, \hat{a}_2^{\dag}\hat{a}_2$ give
the quasiparticle populations of mode 1 and 2 respectively. We
derive the Heisenberg equations for the mode operators by the
relation
\begin{equation}
\dot{a}_i=\frac{i}{\hbar}\left[H,a_i\right].
\end{equation}
To remove the fast oscillation  at the mode frequencies
$\omega_1,\omega_2$ from the operators $\hat{a}_1,\hat{a}_2$ we
introduce the slowly varying operators $\hat{b}_1,\hat{b}_2$:
$\hat{b}_1=\hat{a}_1\frac{1}{\sqrt{N_0}}e^{i\omega_1t},\hat{b}_2=\hat{a}_2\frac{1}{\sqrt{N_0}}e^{i\omega_2t}$,
where the $\sqrt{N_0}$ factor arises because of the different
normalization.

 The Heisenberg equations in terms of these operators are
\begin{eqnarray}
i\hbar\frac{d \hat{b_1}}{dt}=\sqrt{N_0}\hbar\kappa
\hat{b}_1^{\dag}\hat{b}_2 e^{-i\Delta_{12}t}\\ i\hbar
\frac{d\hat{b}_2}{dt}=\frac{1}{2}\sqrt{N_0}\hbar\kappa\hat{b}_1\hat{b}_1
e^{i\Delta_{12}t}. \label{operatorequations}
\end{eqnarray}
Eqs.(\ref{coupledbees}) are obtained by replacing the mode
operators by complex numbers and setting
\begin{equation}
\sqrt{N_0}\hbar\kappa=N_0 g M_{12}.
\end{equation}
However, quantum effects such as squeezing and sub-Poissonian
statistics in the quasi-particle number can only be described by
the operator equations (\ref{operatorequations}) and are lost in
making the classical approximation leading to
Eqs.(\ref{coupledbees}), where operators are replaced by complex
numbers.

\subsection{Squeezing in Parametric Down- and Up-conversion}
In this subsection we want to show  briefly how a quantum
description of the nonlinear processes leads to nonclassical
effects such as squeezing. We apply the well established theory of
nonlinear effects in optics \cite{walls,mandel} to describe the
phononic coupling in a condensate and demonstrate the direct
dependence of squeezing on the coupling matrix element in both,
up- and down-conversion processes. It is difficult to study
squeezing for the full quantum mechanical description of both
modes given by the Hamiltonian (\ref{secondquantham}) as the
operator equations are nonlinear.
 We first investigate down-conversion for resonant coupling and
transform the Hamiltonian (\ref{secondquantham}) into the
interaction picture to obtain
\begin{equation}
H_R =-i\frac{\hbar\kappa}{2}(\hat{b}_1^{\dag
2}\hat{b}_2-\hat{b}_1^2
\hat{b}_2^{\dag}),\label{interactionpicture}
\end{equation}
where $\hat{b}_1,\hat{b}_1^{\dag}$ denote the operators
$\hat{a}_1,\hat{a}_1^{\dag}$ transformed into the interaction
picture. Then the mode operators $\hat{b}_2,\hat{b}_2^{\dag}$ for
mode $2$ are replaced by the c-number $\beta_2$ (we can without
loss of generality assume that $\beta_2$ is real), but we retain
the operators for mode 1. In addition we will treat the mode
amplitude of the upper level as a constant. This assumption does
not account for the depletion of the upper mode and is only valid
for small times when the occupation of the lower level is much
smaller than the occupation of the upper level, i.e. for $N_1\ll
\beta_2^2$. The resulting Hamiltonian is quadratic in
$\hat{b}_1,\hat{b}_1^{\dag}$ and gives linear Heisenberg
equations:
\begin{eqnarray}
\frac{d\hat{b}_1}{dt} &=&[\hat{b}_1,H_R]=\kappa \beta_2
\hat{b}_1^{\dag}\nonumber\\ \frac{d\hat{b}_1^{\dag}}{dt} &=&
[\hat{b}_1^{\dag},H_R]=\kappa \beta_2
\hat{b}_1.\label{linearequations}
\end{eqnarray}

Eqs.(\ref{linearequations}) can be diagonalised by expressing them
in terms of the two quadrature phase amplitudes
$\hat{Q}_x,\hat{Q}_p$ defined as
\begin{eqnarray}
\hat{Q}_x &=&\hat{b}_1+\hat{b}_1^{\dag}\nonumber\\ \hat{Q}_p &
=&\frac{\hat{b}_1-\hat{b}_1^{\dag}}{i}.
\end{eqnarray}
Simple integration yields $\hat{Q}_x(t)=e^{\kappa
\beta_2t}\hat{Q}_x(0),\hspace{0.2em}\hat{Q}_p(t)=e^{-\kappa
\beta_2t}\hat{Q}_p(0)$. These solutions can be used to calculate
the evolution of $\hat{b}_1,\hat{b}_1^{\dag}$ and the evolution
for the number of down-converted quasi-particles, $N_1$, for which
we find (assuming mode $2$ was initially in a vacuum state
$|0\rangle$):
\begin{equation}
N_1=\langle 0| \hat{b}_1^{\dag}\hat{b}_1|0 \rangle =
\sinh^2(\kappa \beta_2t). \label{downconvertednumber}
\end{equation}

Eq.(\ref{downconvertednumber}) shows that in this quantum model
downconversion occurs even for zero initial population in mode
$1$. This is in contrast to the result of the semiclassical model
discussed above. The evolution of the variances in the quadrature
operators is found to be:
\begin{eqnarray}
\left[\Delta\hat{Q}_x(t)\right]^2&=&e^{2\kappa \beta_2 t}
\left[\Delta\hat{Q}_x(0)\right]^2\nonumber\\
\left[\Delta\hat{Q}_p(t)\right]^2&=&e^{-2\kappa \beta_2
t}\left[\Delta\hat{Q}_p(0)\right]^2. \label{uncertainties}
\end{eqnarray}
This clearly demonstrates the squeezing in the $\hat{Q}_p$
quadrature component. However, it is important to keep in mind
that Eqs.(\ref{uncertainties}) are only valid for short times
before the assumption that the upper mode is not depleted breaks
down. A possible way to avoid depletion of the upper mode is to
keep exciting it.  The standard way to do this is to mechanically
force the condensate into oscillation at the frequency and
geometry corresponding to the upper mode.

Similarly, squeezing occurs during SHG where two phonons from the
lower mode are converted into one phonon of the upper mode. In
this case we cannot replace the lower mode by a c-number as we
have done for our investigation of downconversion. We can again
try to find an approximate solution valid only for small times.
This is demonstrated in \cite{mandel}, where a Taylor series
expansion is used to describe the time evolution of the mode
operators. We assume that mode $1$ is initially in a coherent
state defined by $b_1|\beta_1\rangle=\beta_1|\beta_1\rangle$ with
$\beta_1=|\beta_1|e^{i|\Phi}$ and mode $2$ is in the vacuum state.
The result for the squeezing of the quadrature $\hat{Q}_x$ (of
mode 1) to second order in time is then given by:
\begin{equation}
\left[\Delta\hat{Q}_x(t)\right]^2=1-\frac{1}{2}\kappa^2
t^2|\beta_1|^2\cos(2\Phi) +O(gt)^3, \label{SHGsqueezing}
\end{equation}
This result only holds for small times for which
$\frac{1}{4}|\beta_1|^2\kappa^2t^2\ll 1$. However, we can see how
in both cases, down- and up-conversion, the squeezing of the
quadrature components can be directly related to the nonlinear
coupling strength.

\section{Calculating the nonlinear coupling rate}
\label{calcnonlincouprate} We will now calculate the nonlinear
coupling rates between two condensate excitations and compare them
to those from recent experiments. For convenience we take linear
combinations of the normalized quasiparticle wavefunctions to give
the new functions $\tilde{f}_i^+,\tilde{f}_i^-$ defined by:
\begin{eqnarray}
\tilde{f}_i^+ &=&\tilde{u}_i+\tilde{v}_i=f_i^+ \\ \tilde{f}_i^- &
=& \tilde{u}_i-\tilde{v}_i=f_i^- -2c_i\psi_g.
\end{eqnarray}
Written in terms of these functions the matrix element in
Eq.(\ref{matrixelement}) has the form
\begin{equation}
M_{12}=2\int d^3{\bf r}\hspace{0.5em}\psi_g\left\{
\frac{1}{2}\tilde{f}_1^{+*}\left(\tilde{f}_1^{+*}\tilde{f}_2^{+}+\tilde{f}_1^{-*}\tilde{f}_2^{-}\right)+
\frac{1}{4}\tilde{f}_2^{+}\left(\tilde{f}_1^{+*}\tilde{f}_1^{+*}-\tilde{f}_1^{-*}\tilde{f}_1^{-*}\right)\right\},
\end{equation}
where we assumed that $\psi_g$ is real. Alternatively, $M_{12}$
can be written in terms of untilded functions ($f_i^+,f_i^-$) as
the sum of two parts, $M_{12}=M_{12}^{(1)}+M_{12}^{(2)}$, which
are defined as follows:
\begin{eqnarray}
 M_{12}^{(1)} &=& 2\int d^3{\bf r}\hspace{0.5em}\psi_g\left\{\frac{1}{2}f_1^{+*}\left(f_1^{+*}+f_2^{+}+f_1^{-*}f_2^{-}\right)+
\frac{1}{4}f_2^{+}\left(f_1^{+*}f_1^{+*}-f_1^{-*}f_1^{-*}\right)\right\},\label{firstpart}\\
M_{12}^{(2)} &=& 2\int d^3{\bf r}\hspace{0.5em}\psi_g\left\{
f_1^{+*}\left(2c_1^*c_2\psi_g^2-c_1^*\psi_g f_2^- -c_2\psi_g
f_1^{-*}\right)+f_2^+\left(c_1^*\psi_g
f_1^{-*}-c_1^{*2}\psi_g^2\right)\right\}.\label{secondpart}
\end{eqnarray}
$M_{12}^{(2)}$ is zero if neither of the quasiparticle
wavefunctions has any overlap with the condensate ground state.
 We can now use the functions $f_i^+,f_i^-$ which we found in section \ref{hydrosolsection}.
 We can write in general:
\begin{equation}
\begin{array} {ll}
f_1^+=A_1\epsilon_1\xi(1-y^2)^{-1/2}W_1, &
f_1^-=A_1(1-y^2)^{1/2}W_1, \\
f_2^+=A_2\epsilon_2\xi(1-y^2)^{-1/2}W_2, &
f_2^-=A_2(1-y^2)^{1/2}W_2,
\end{array}
\label{fdefinitions}
\end{equation}
where the $A_1,A_2$ are normalization constants determined from
the normalization condition (\ref{orthogonality}) and
$W_1(y_1,y_2,y_3),W_2(y_1,y_2,y_3)$ are solutions to
Eq.(\ref{Gequation}) for mode $1$ and $2$ respectively.
Substituting these expressions into Eq.(\ref{firstpart}) gives:
\begin{equation}
M_{12}^{(1)}=(\sqrt{n_0/N_0}A_1^2 A_2\xi l_c^3/2)\int d^3{\bf y}
W_1^2
W_2\left[3\epsilon_1^2\epsilon_2\xi^2(1-y^2)^{-1}-\Delta\epsilon_{12}(1-y^2)\right].
\label{nonorthogonalpart}
\end{equation}

The first term in the above integral proportional to
$(1-y^2)^{-1}$ diverges at the condensate boundary ($y=1$), but it
must be dropped as it scales proportional to $\xi^2$ and in the
derivation of the quasiparticle wavefunctions we omitted terms
proportional to $\xi^2$ in the governing equations (hydrodynamic
approximation) to obtain Eq.(\ref{Gequation}) for
$W(y_1,y_2,y_3)$. We will see later in this paper that this is
fully justified by comparison with exact numerical calculations.
Note that the second term equals zero if the detuning
$\Delta\epsilon_{12}=\epsilon_2-2\epsilon_1$ is zero.

\subsection{Coupling between two even parity quadrupolar
excitations} So far we have made no assumptions about the geometry
of the two modes that are coupled and
Eqs.(\ref{firstpart}-\ref{secondpart}) are valid for any pair of
modes. We now focus on the quadrupolar modes of a triaxial trap
and investigate the coupling between any two diagonal modes, for
which the function $W(y_1,y_2,y_3)$ is represented by the
polynomial given in Eq.(\ref{polynomialW}). We can calculate the
matrix element $M_{12}$ from
Eqs.(\ref{secondpart},\ref{nonorthogonalpart}). An explicit
expression for $M_{12}$ and its derivation is given in Appendix A.
It is important to note that for on-resonant coupling between the
two modes ($\omega_2=2\omega_1$) the matrix element simplifies
considerably. In that case we can give a simple expression for the
radial integrand of the matrix element in terms of the
dimensionless position $y$, where the condensate boundaries are
given by $y=1$:
\begin{equation}
M_{12}=-2\sqrt{\frac{n_0}{N_0}}A_2\int_0^1
y^6(1-y^2)dy=-2\sqrt{\frac{n_0}{N_0}}A_2\frac{2}{63},
\label{mainresult}
\end{equation}
where $A_2$ denotes the normalization constant for the
wavefunction of mode $2$, given by
Eq.(\ref{normalizationconstant}). It is important to note that
Eq.(\ref{mainresult}) describes the resonant coupling between any
two diagonal quadrupole modes in a triaxial trap. This allows a
quick and easy calculation of the coupling strength, coupling
rates and squeezing effects associated with the nonlinear process.
We can see from the radial integrand of Eq.(\ref{mainresult}) that
the coupling is strongest in the outer regions of the condensate
and reaches a maximum at a distance of $0.87$ $l_c$ from the
center. In this region the condensate is still well described by
the hydrodynamic approximation which only breaks down at a
distance of order the healing length from the condensate
boundaries. The healing length in our recent experiments
\cite{hechenblaikner,hodby} was about $0.05$ $l_c$.

\subsection{Coupling between two modes in a spherical trap}
Now we want to give a quantitative comparison between the
solutions we found to the coupling matrix element in a
hydrodynamic approach and the exact solutions calculated from the
numerical solutions to the BdG equations. To facilitate the
numerical calculations we will look at the coupling between two
quadrupolar modes in a spherical trap of frequency $\omega$, where
the total angular momentum $l$ and the azimuthal angular momentum
$m$ are good quantum numbers. The two modes are the $l=2,m=0$ mode
and the $l=0,m=0$ mode with frequencies of $\sqrt{2}\omega$ and
$\sqrt{5}\omega$ respectively. In a trap with only axial symmetry
these two modes get mixed and become the $m=0$ low-lying and the
$m=0$ high-lying (breathing) mode. Their quasiparticle
wavefunctions can be presented by the ansatz (\ref{fdefinitions})
where $W_1=(3\cos^2\theta-1)y^2$ and $W_2=(1-5y^2/3)$. These are
the solutions to the hypergeometric differential equations
discussed in section \ref{hydrosolsection}. The normalization
constants are $A_1=\sqrt{35/16\pi l_c^3\epsilon_1\xi}$ and
$A_2=(3/2)\sqrt{7/4\pi l_c^3\epsilon_2\xi}$.

\begin{figure}
\begin{center}\mbox{ \epsfxsize 3.2 in\epsfbox{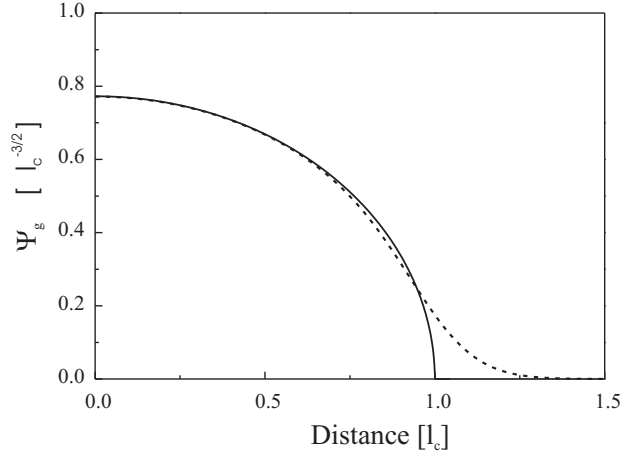}}\end{center}
\caption{The wavefunctions for the ground state in the
hydrodynamic limit (solid line) and the exact solution to the GPE
(dotted line) plotted in units of $1/\sqrt{l_c^3}$ against the
distance from the center of the trap in units of the
characteristic length $l_c$. The trap is spherical with $1.5\times
10^4$ atoms and a frequency $\omega=120$ Hz . The healing length
$\xi=0.05 l_c$} \label{wavefun}
\end{figure}

\begin{figure}
\begin{center}\mbox{ \epsfxsize 3.2 in\epsfbox{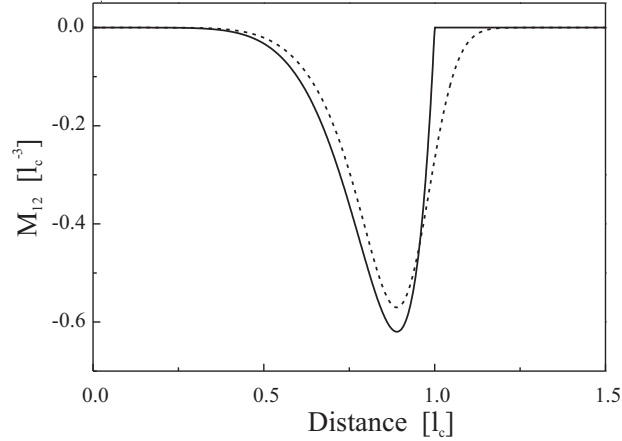}}\end{center}
\caption{The radial integrands for the matrix-elements $M_{12}$ in
the hydrodynamic limit (solid line) and for the exact numerical
calculation (dotted line) plotted in units of $l_c^{-3}$ against
the distance from the condensate center in units of $l_c$. The
coupling is between the $l=0,m=0$ and the $l=2,m=0$ mode for the
same trap conditions as in Fig.(\ref{wavefun}).}
\label{matrixplot}
\end{figure}

One can see from Eq.(\ref{secondpart}) that in $M_{12}^{(2)}$ all
terms containing the constant $c_1$ are zero, because the overlap
between mode 1, which is proportional to $Y_2^0$, and the ground
state, which is proportional to $Y_0^0$, is zero. The only
remaining term is proportional to $c_2$ and it turns out to be
$-2\sqrt{n_0/N_0}A_2 y^6(1-y^2)$, which is exactly the expression
we found for the resonant case ($\Delta_{12}=0$) in a general
triaxial trap (see Eq.(\ref{mainresult})). But for these two modes
$\Delta_{12}\neq 0$ and we have to consider the contribution from
$M_{12}^{(1)}$ as well so that we obtain
\begin{equation}
M_{12}=-2\sqrt{\frac{n_0}{N_0}}A_2\int_0^1
\left[\frac{7}{4\epsilon_1}\Delta_{12}(1-y^2)(1-5/3y^2)+(1-y^2)\right]y^6dy
\end{equation}
The wave functions for the ground state and the coupling matrix
elements $M_{12}$ are plotted in Fig.(\ref{wavefun}) and
Fig.(\ref{matrixplot}) respectively. It is important to note that
for the resonant case the only contribution comes from
Eq.(\ref{mainresult}). Also, for the not quite resonant case
displayed in Fig.(\ref{matrixplot}) the integrand is dominated by
this contribution. This shows that the coupling between different
quasi-particle excitations predominantly takes place in the
boundary region of the condensate and an explicit analytic
expression for the spatial probability of the nonlinear process is
given by Eq.(\ref{mainresult}).

The integrated values of $M_{12}$ in the hydrodynamic
approximation and for the exact numerical calculation  are
$-0.161\hspace{0.5em} l_c^{-3}$ and $-0.157\hspace{0.5em}
l_c^{-3}$ respectively. The error of the approximate analytical
result with respect to the exact numerical calculation is in the
order of $\xi^2$, as we expect. The good agreement between the
two, even for a relatively small number of atoms, justifies the
hydrodynamic approximations made in calculating the quasi particle
wavefunctions and the coupling matrix element.

\subsection{Coupling between two even parity modes in a TOP-trap}
\begin{figure}
\begin{center}\mbox{ \epsfxsize 3.2 in\epsfbox{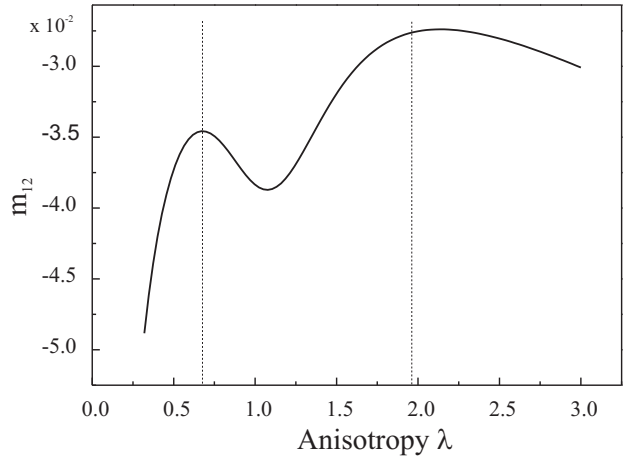}}\end{center}
\caption{The quantity $m_{12}=M_{12}l_c^3\sqrt{\xi}$ is plotted
against the trap anisotropy $\lambda$. The two vertical lines show
where there is resonant nonlinear coupling between the $m=0$
low-lying and the $m=0$ high-lying modes. The resonances are
determined by the matching of mode frequencies
($2\omega_1=\omega_2$) and they are located at $\lambda=0.68$ and
$\lambda=1.95$.} \label{topcoupling}
\end{figure}

In our experiments we use a TOP-trap which is axially symmetric
and has an anisotropy defined by the parameter
$\lambda=\omega_z/\omega_r$, where $\omega_z$ and $\omega_r$ are
the axial and radial trap frequencies respectively. In a recent
experiment we studied the coupling between the $m=0$ low-lying and
the $m=0$ high-lying mode (which arise from the $l=0,m=0$ and the
$l=2,m=0$ mode of the spherical trap when $\lambda\neq 1$) . We
can use the formula for $M_{12}$ for the general triaxial trap,
given by Eqs.(\ref{triax1},\ref{triax2}) in the Appendix, to
calculate the matrix element for any trap geometry. The result is
shown in Fig.(\ref{topcoupling}) in terms of the dimensionless
quantity $m_{12}=M_{12}l_c^3\sqrt{\xi}$. The dependence of
$M_{12}$ on number and mean-frequency is contained in
$l_c^{-3}\xi^{-1/2}$ and thus $m_{12}$ only depends on the mode
geometry, i.e. it is only a function of $\lambda$. In order to get
$M_{12}$ for a specific trap, one has to read the dimensionless
matrix element $m_{12}$ from Fig.(\ref{topcoupling}) and multiply
it by $l_c^{-3}\xi^{-1/2}$.

From Eq.(\ref{topequation}) one can find two values for the
anisotropy $\lambda$ where the two $m=0$ modes are resonantly
coupled. These resonances are at $\lambda=0.68,1.95$ with
$|m_{12}|=0.038,0.028$ respectively.
 The characteristic coupling time is given by Eq.(\ref{transfertime})
which can be written as follows:
\begin{equation}
T=\frac{15\sqrt{2} \hbar \sqrt{\xi}}{8\pi\mu m_{12}b_1(0)}\propto
(N^{-3/5}\bar{\omega}^{-13/10}), \label{transfernew}
\end{equation}
We can calculate the coupling rate for the parameters of our
second harmonic generation experiment \cite{hechenblaikner}, which
were: $\lambda=1.95,N_0=1.5\times 10^4,\omega_{r}\simeq 120$ Hz.
Inserting the relevant quantities into Eq.(\ref{transfernew}) we
obtain a transfer time of $5.7$ ms and $3.7$ ms for small initial
excited populations of $|b_1(0)|^2=0.02$ and $|b_1(0)|^2=0.05$
respectively. For stronger initial excitation the transfer times
would be even smaller. One can see from Eq.(\ref{transfernew})
that for larger atom number and stiffer traps the coupling times
become smaller as well. This shows that the process of coupling
the two quadrupole modes happens on a timescale of a few
milliseconds. It is consistent with our experimental observation
\cite{hechenblaikner} where the second-harmonic was observed as
soon as the driving of the fundamental finished (excitation time
of about $30$ ms).

\subsection{Coupling between a scissors mode and another
quadrupolar excitation}
 Now we will discuss coupling between the off-diagonal
quadrupolar excitations. We already showed in section
\ref{hydrosolsection} that these modes have odd parity and are
characterized by $W\propto y_iy_j, i\neq j$. Thus their overlap
with the condensate ground state is zero and the only contribution
to the coupling matrix element comes from $M_{12}^{(1)}$ given in
Eq.(\ref{nonorthogonalpart}). But this integral is zero because
the product $W_1^2W_2$ is odd. So the total matrix element
$M_{12}$ equals zero and there is no direct coupling via a second
harmonic process between two scissors modes. Our recent
experimental observation of a downconversion process between two
scissors modes must therefore have a more sophisticated
interpretation than the conversion of one quantum of the higher
mode into two quanta of the lower scissors mode.

 Similarly, there
is no coupling between a higher-lying scissors mode and a
lower-lying diagonal quadrupole mode. In this case $M_{12}^{(1)}$
of Eq.(\ref{nonorthogonalpart}) is zero beacuse $W_2$ is odd.
$M_{12}^{(2)}$ of Eq.(\ref{secondpart}) is also zero because
$c_2=0$ and $f_2^{\pm}$ are odd. Thus the total matrix element is
zero. However, there is second harmonic coupling from a lower
scissors mode to a higher lying diagonal quadrupolar mode. For
resonant coupling two quanta of the scissors mode are converted
into one quantum of the higher lying even parity mode and one
finds that the matrix element is again given by
expression(\ref{mainresult}).

\section{conclusion}
Starting from the NLSE we have derived a simple model describing
the nonlinear coupling between two modes. We have demonstrated a
way how to calculate analytically the matrix element governing the
coupling process. We then focused on the quadrupole excitations
and found that all resonant ($\omega_2=2\omega_1$) direct coupling
processes between the six quadrupole modes are described by the
expression in Eq.(\ref{mainresult}) unless they are forbidden
($M_{12}=0$). All second harmonic processes involving odd parity
(scissors) modes are forbidden except for upconversion from a
scissors mode to a higher lying even parity mode. It is possible
to show that there are other allowed nonlinear processes involving
all three of the scissors modes. This gives rise to nondegenerate
parametric amplification and multimode squeezing. Full details and
the derivation of an effective Hamiltonian describing all allowed
nonlinear processes between the quadrupole modes (not just the
second harmonic generation described here) will be given in a
future publication.

\acknowledgements  We would like to thank H. Nilsen, H. Ritsch, C.
Lamprecht, J. Dunningham, S. Choi and K. Burnett for fruitful
discussions. We acknowledge support from the EPSRC, St. John's
College, Oxford (G.H.), Trinity College, Oxford (S.A.M.)  and the
EC (O.M.).

\begin{appendix}

\section{matrix elements for the diagonal quadrupolar modes}
We want to calculate the coupling matrix element for any two
diagonal quadrupole modes in a triaxial trap for which the
function $W(y_1,y_2,y_3)$ is represented by a polynomial as given
in Eq.(\ref{polynomialW}). It is useful to derive a number of
relations for the polynomial coefficients which allow us to
simplify the expressions for normalization constants, overlap
coefficients and the coupling matrix element.

The lowest mode of Eq. (\ref{quantansatz}) is the so called
Goldstone mode with $\omega_0=0, u_0({\bf r})=\Psi_g({\bf r})$ and
$v_0({\bf r})=-\Psi_g^*({\bf r})$, which arises from the $U(1)$
symmetry breaking. For this particular mode the orthogonality and
symmetry relations of Eqs.(\ref{orthogonality}) take the form
\begin{equation}
\int d^3{\bf r} \left(\Psi_g^*u_i+\Psi_g v_i\right)=0.
\label{fpluspsig}
\end{equation}
Eq.(\ref{fpluspsig}) implies that $\int d^3{\bf r} \psi_g f^+=0$.
Substituting (\ref{fdefinitions}) into this equation and
integrating over the angles and radial coordinate gives:
\begin{equation}
 \sum_j\bar{b}_j=-5 \label{sumfive}
\end{equation}
The characteristic polynomials for the quadrupole modes are real
and thus $u_i,v_i$ can be taken as real which allows us to derive
from the orthogonality relations (\ref{orthogonality}):
\begin{equation}
\int d^3{\bf x} f^+_i f^-_j=\delta_{ij}. \label{neworthogonality}
\end{equation}
 If we now insert two
different polynomials corresponding to two different solutions for
$f^+_i,f^-_j$ into Eq.(\ref{neworthogonality}) we obtain the
relation $\int d^3{\bf y} W_1 W_2=0$ and from that:
\begin{eqnarray}
\sum_i\bar{b}_i\bar{d}_i=5\\
\bar{b}_1\bar{d}_2+\bar{b}_2\bar{d}_1+\bar{b}_1\bar{d}_3+\bar{b}_3\bar{d}_1+\bar{b}_2\bar{d}_3+\bar{b}_3\bar{d}_2=20,
\end{eqnarray}
where the $\bar{b}_i$'s denote the polynomial coefficients of mode
$1$ and the $\bar{d}_i$'s the polynomial coefficients of mode $2$.
These coefficients are found from
Eqs.(\ref{Sequations},\ref{normequation}) in section
\ref{hydrosolsection}. We also have to calculate the overlap
coefficients $c_i$ between the condensate and the quasiparticle
wavefunctions:
\begin{eqnarray}
c_i=\int d^3{\bf r} \psi_g^*u_i=\frac{1}{2}\int d^3{\bf r}
\psi_g^* (f_i^+ + f_i^-)=\frac{1}{2}\int d^3{\bf r} f_i^-.
\label{overlapcoefficient}
\end{eqnarray}
After substituting Eqs.(\ref{fdefinitions}) into
Eq.(\ref{overlapcoefficient}) we obtain
\begin{equation}
c_i=\frac{1}{2}A_i l_c^3
\sqrt{\frac{n_0}{N_0}}\frac{8\pi}{15}\left(1+\frac{1}{7}\sum_j
\bar{b}_j\right)=\frac{1}{7}A_i \sqrt{\frac{N_0}{n_0}},
\end{equation}
where we used (\ref{sumfive}) and $N_0=n_0 l_c^38\pi/15$ (Thomas
Fermi relation for $\mu(N_0)$) for the last step. We also have to
calculate the normalization amplitudes for these modes and obtain
from Eqs.(\ref{neworthogonality}):
\begin{equation}
A_i=\left(\epsilon_i\xi l_c^3\frac{4\pi}{105}\left[3\sum_j
\bar{b}_j^2+2(\bar{b}_1\bar{b}_2+\bar{b}_1\bar{b}_3+\bar{b}_2\bar{b}_3)-35\right]\right)^{-1/2}.
\label{normalizationconstant}
\end{equation}

We can now calculate the coupling matrix element from
Eqs.(\ref{secondpart},\ref{nonorthogonalpart}). We shall introduce
a constant $R_{12}$ defined as follows:
\begin{equation}
R_{12}=\bigg(\sqrt{\frac{n_0}{N_0}} A_1^2 A_2\xi
l_c^3/2\bigg)\frac{4\pi}{105}.
\end{equation}
The first part of the matrix element $M_{12}^{(1)}$ is then:
\begin{eqnarray}
M_{12}^{(1)} &=& -R_{12}\int_0^1 dy \bigg\{y^6\bigg[15\sum_i
\bar{b}_i^2\bar{d}_i+3(2\bar{b}_1\bar{b}_2\bar{d}_1+\bar{b}_1^2\bar{d}_2+2\bar{b}_1\bar{b}_3\bar{d}_1+
\bar{b}_1^2\bar{d}_3+2\bar{b}_1\bar{b}_2\bar{d}_2+\bar{b}_2^2\bar{d}_1+\nonumber\\
&&2\bar{b}_1\bar{b}_3\bar{d}_3+\bar{b}_3^2\bar{d}_1+
2\bar{b}_2\bar{b}_3\bar{d}_2+\bar{b}_2^2\bar{d}_3+2\bar{b}_2\bar{b}_3\bar{d}_3+\bar{b}_3^2\bar{d}_2)+
2\bar{b}_1\bar{b}_2\bar{d}_3+2\bar{b}_1\bar{b}_3\bar{d}_2+2\bar{b}_2\bar{b}_3\bar{d}_1
\bigg]+\nonumber\\
&&y^4\bigg[490+21\sum_i\bar{b}_i^2+7(2\bar{b}_1\bar{b}_2+2\bar{b}_2\bar{b}_3+2\bar{b}_3\bar{b}_1)\bigg]-
525 y^2+105\bigg\} y^2 \Delta\epsilon_{12}(1-y^2)dy.
\label{triax1}
\end{eqnarray}
The second part of the matrix element is given in
Eq.(\ref{triax2}). Note that this part arises due to the finite
overlap between the condensate ground state and the untilded
Bogoliubov wavefunctions $u_i,v_i$.
\begin{eqnarray}
M_{12}^{(2)}&=& 4R_{12}\int_0^1\Delta\epsilon_{12}\left\{35 y^4
-\frac{325}{7} y^2 +\frac{90}{7}\right\}y^2(1-y^2) dy\nonumber\\
&& -2\sqrt{\frac{n_0}{N_0}}A_2\int_0^1 y^6 (1-y^2).
dy\label{triax2}
\end{eqnarray}
We note that for resonant interaction
$\Delta\epsilon_{12}=\epsilon_2-2\epsilon_1=0$, so that
$M_{12}^{(1)}=0$ and the only term surviving in $M_{12}^{(2)}$ is
$-2\sqrt{n_0/N_0}A_2\int_0^1 y^6 (1-y^2) dy$.
\end{appendix}





\end{document}